\documentclass[12pt]{article}

\usepackage[a4paper,text={16.3cm,22cm}]{geometry}
\usepackage{amsmath,amsfonts,mathtools,braket,slashed,amssymb,bm,bbm,graphicx}
\usepackage[table,dvipsnames]{xcolor}
    \definecolor{Lightgray}{gray}{0.93}
\usepackage[labelfont=bf,width=0.95\textwidth]{caption}
\usepackage{cite}
\usepackage[
    bookmarksnumbered=true,
    urlcolor=blue,
    linkbordercolor=red,
    citebordercolor=green,
    bookmarksopen=true
    ]{hyperref}

\allowdisplaybreaks
\numberwithin{equation}{section}

\newcommand{\spac}{{\hspace{0.3mm}}}
\newcommand*\leftlap[3][\,]{#1\hphantom{#2}\mathllap{#3}}
\newcommand*\rightlap[2]{\mathrlap{#2}\hphantom{#1}}

\title{Super-Leading Logarithms in \texorpdfstring{$pp\to2$}{pp->2} Jets}

\begin{document}

\begin{titlepage}

\begin{flushright}
{\small
MITP-24-082\\
TUM-HEP-1536/24\\
November 19, 2024
}
\end{flushright}

\makeatletter
\vskip0.8cm
\pdfbookmark[0]{\@title}{title}
\begin{center}
{\Large\bf\boldmath \@title}
\end{center}
\makeatother

\vspace{0.5cm}
\begin{center}
\textsc{Thomas Becher,$^a$ Patrick Hager,$^b$ Giuliano Martinelli,$^a$\\
Matthias Neubert,$^{a,b,c}$ Dominik Schwienbacher$^a$ and Michel Stillger$^d$}\\[6mm]
    
\textsl{${}^a$Albert Einstein Center for Fundamental Physics, Institut für Theoretische Physik\\
Universität Bern, Sidlerstrasse 5, CH-3012 Bern, Switzerland\\[0.3cm]
${}^b$PRISMA$^+$ Cluster of Excellence \& Mainz Institute for Theoretical Physics\\
Johannes Gutenberg University, Staudingerweg 9, D-55128 Mainz, Germany\\[0.3cm]
${}^c$Department of Physics \& LEPP, Cornell University, Ithaca, NY 14853, U.S.A.\\[0.3cm]
${}^d$Physik Department T31, Technische Universit\"at M\"unchen\\
James-Franck-Straße 1, D-85748 Garching, Germany}
\end{center}

\vspace{0.6cm}
\pdfbookmark[1]{Abstract}{abstract}
\begin{abstract}
\noindent 
Jet observables at hadron colliders feature ``super-leading'' logarithms, double-loga\-rithmic corrections resulting from a breakdown of color coherence due to complex phases in hard-scattering amplitudes. While these effects only arise in high orders of perturbation theory and are suppressed in the large-$N_c$ limit, they formally constitute leading logarithmic corrections to the cross sections. We present the first analysis of the corresponding contributions to a hadronic cross section, including all partonic channels and interference effects. Interestingly, some interference terms in partonic $q\bar q\to q\bar q$ scattering are only linearly suppressed in $1/N_c$. Our results for the $pp\to 2$\,jets gap-between-jets cross section demonstrate the numerical importance of super-leading logarithms for small values of the veto scale $Q_0$, showing that these contributions should be accounted for in precision studies of such observables.
\end{abstract}

\vfill\noindent\rule{0.4\columnwidth}{0.4pt}\\
\hspace*{2ex} {\small \textit{E-mail:} \href{mailto:becher@itp.unibe.ch}{becher@itp.unibe.ch}, \href{mailto:pahager@uni-mainz.de}{pahager@uni-mainz.de}, \href{mailto:giuliano.martinelli@students.unibe.ch}{giuliano.martinelli@students.unibe.ch},\\
\hspace*{2ex} \phantom{E-mail: } \href{mailto:matthias.neubert@uni-mainz.de}{matthias.neubert@uni-mainz.de}, \href{mailto:dominik.schwienbacher@unibe.ch}{dominik.schwienbacher@unibe.ch}, \href{mailto:m.stillger@tum.de}{m.stillger@tum.de}}

\end{titlepage}

\section{Introduction}

The increasing precision of experimental measurements at the CERN Large Hadron Collider (LHC) challenges theoretical predictions, enforcing the need for higher-order perturbative computations and resummations of logarithmically enhanced corrections. For jet observables, in particular, the latter feature an intricate structure, arising due to non-global logarithms (NGLs)~\cite{Dasgupta:2001sh}, which originate from secondary soft emissions, and super-leading logarithms (SLLs)~\cite{Forshaw:2006fk,Forshaw:2008cq,Keates:2009dn}, which are related to the breaking of collinear factorization in space-like collinear splittings~\cite{Catani:2011st,Forshaw:2012bi,Schwartz:2017nmr,Forshaw:2021fxs}. Whereas NGLs are present also for $e^+e^-$ colliders, SLLs appear at hadron colliders only, where the scattered partons carry color.

While traditional parton showers compute some logarithmically-enhanced contributions, their systematic accuracy is no longer sufficient for precision physics. There has recently been much progress in systematically improving parton showers by including higher-order logarithmic corrections~\cite{Dasgupta:2020fwr,Forshaw:2020wrq,vanBeekveld:2022zhl,Herren:2022jej,FerrarioRavasio:2023kyg,Preuss:2024vyu}, and showers with resummation up to next-to-next-to-leading-logarithmic accuracy are now becoming a reality~\cite{vanBeekveld:2024wws}. Parton showers commonly work in the large-$N_c$ limit. Corrections to this limit are often moderate and can be approximated by adjusting color factors, see e.g.\ \cite{Hamilton:2020rcu,Holguin:2020joq}. SLLs are a remarkable exception, as they constitute leading double-logarithmic effects but vanish in the large-$N_c$ limit. They are therefore missed by traditional parton showers. To capture their effects, one needs an amplitude-level parton shower with full color information. Such showers are currently being developed~\cite{AngelesMartinez:2018cfz,Nagy:2019pjp,Forshaw:2019ver,DeAngelis:2020rvq}, but a dedicated study of SLLs in these frameworks is not yet available.

The simplest observables featuring SLLs are gap-between-jets cross sections. Here, one imposes a veto on radiation above a scale $Q_0$ inside a gap region, which lies outside the hard jets with transverse momentum of order $Q$, resulting in large logarithmic corrections in $L=\ln(Q/Q_0)\gg 1$. In practice, one vetoes jets with transverse momentum above $Q_0$ inside the gap region~\cite{ATLAS:2011yyh}. The formalism developed in~\cite{Becher:2021zkk,Becher:2023mtx,Boer:2024hzh} allows for a resummation of these logarithms using renormalization-group (RG) techniques. Before resummation, the all-order series of SLLs to a non-global partonic cross section takes the form
\begin{equation}\label{eq:SLLgeneric}
   \hat\sigma_{2\to M}^{\rm SLL}(Q_0) 
   = \hat\sigma_{2\to M}\,\frac{\alpha_s\spac L}{\pi\spac N_c} \left( \frac{N_c\spac\alpha_s}{\pi}\,i \pi L \right)^2
    \sum_{n=0}^\infty\,c_n \left( \frac{N_c\spac\alpha_s}{\pi}\,L^2 \right)^{n} ,
\end{equation}
with $\hat\sigma_{2\to M}$ denoting the partonic cross section in the Born approximation. The terms under the sum have alternating signs and constitute the double-logarithmic corrections. The prefactor in parenthesis contains two complex phases $i\pi$ resulting from gluon exchanges between the two colliding partons, which break color coherence and are responsible for the appearance of the SLLs. The factor $\pi^2$ implies a further numerical enhancement of these effects. The remaining prefactor contains $\alpha_s/N_c$ rather than $N_c\spac\alpha_s$, indicating that the SLLs are a color-suppressed effect. From now on, the phrase ``SLL contribution'' will always refer to the entire series in \eqref{eq:SLLgeneric}, including the three-loop contribution (the term with $n=0$), even though the first double logarithm arises at four-loop order ($n\ge 1$). All contributions are due to Glauber phases and thus have the same physical origin.

At the parton level sizable contributions, both positive and negative, of up to $\mathcal{O}(10\%)$ were found for values of $Q_0\sim 20$\,GeV~\cite{Becher:2021zkk,Becher:2023mtx,Boer:2024hzh}, but it remained an open question if there are cancellations in the sum over partonic subprocesses contributing to a hadronic cross section. In the present paper, we perform a full analysis for the $pp\to 2$\,jets process with a gap between the jets, combining all relevant partonic channels, and including interference terms in the squared partonic scattering amplitudes. This provides the first estimate of the numerical impact of the SLLs contribution on a physical cross section, demonstrating that the effects are of similar size to the ones obtained on the parton level.

\section{Formalism}

Building on the factorization theorem for gap-between-jets cross sections at hadron colliders~\cite{Balsiger:2018ezi,Becher:2021zkk,Becher:2023mtx}
\begin{equation} \label{eq:factorization_formula}
   \sigma_{2\to M}(Q_0) 
   = \int\!d\xi_1 \int\!d\xi_2\,\sum_{m=2+M}^\infty \big\langle 
    \bm{\mathcal{H}}_m(\{\underline{n}\},s,\xi_1,\xi_2,\mu) \otimes 
    \bm{\mathcal{W}}_m(\{\underline{n}\},Q_0,\xi_1,\xi_2,\mu) \big\rangle \,,
\end{equation}
a formalism to resum large logarithms was developed in~\cite{Becher:2021zkk,Becher:2023mtx,Boer:2024hzh}. The hard functions $\bm{\mathcal{H}}_m$ describe the partonic hard-scattering process of $m$ partons and capture the physics at a high scale $\mu_h\sim\sqrt{\hat{s}}$ of order the partonic center-of-mass energy. The low-energy matrix elements $\bm{\mathcal{W}}_m$ account for dynamics at the veto scale $\mu_s\sim Q_0$, as well as non-perturbative physics at the scale $\Lambda_{\rm QCD}$. Their product is integrated over the momentum fractions $\xi_i$ of the incoming partons as well as over the directions ($n_3,\dots,n_m$) of the final-state partons. These angular integrations are represented by the $\otimes$ symbol. Finally, $\langle\ldots\rangle$ indicates a trace in color-helicity space. More details on the factorization theorem can be found in~\cite{Becher:2023mtx}.

To resum the SLLs to all orders in perturbation theory, one evaluates relation~\eqref{eq:factorization_formula} at the low scale $\mu_s$. The hard functions are evolved from their characteristic scale $\mu_h$ down to $\mu_s$ by solving their RG equation. At the scale $\mu_s$, the low-energy matrix elements have the simple form
\begin{equation} \label{eq:low_energy_ME}
   \bm{\mathcal{W}}_m(\{\underline{n}\},Q_0,\xi_1,\xi_2,\mu_s) 
   = f_1(\xi_1,\mu_s)\,f_2(\xi_2,\mu_s)\,\bm{1} + \mathcal{O}(\alpha_s) \,,
\end{equation}
where $f_i(\xi_i,\mu_s)$ denotes the parton distribution function (PDF) for parton $i$ of the proton. For a detailed study of $\bm{\mathcal{W}}_m$, we refer the interested reader to~\cite{Becher:2024kmk}.

Using a compact matrix notation, the solution of the RG equation for the hard functions can be expressed as a path-ordered exponential
\begin{equation} \label{eq:H_RGE}
    \bm{\mathcal{H}}(\mu_s) 
    = \bm{\mathcal{H}}(\mu_h) \star \mathbf{P} 
     \exp\biggl[ \int_{\mu_s}^{\mu_h}\frac{d\mu}{\mu}\,\bm{\Gamma}^H(\mu) \biggr] \,,
\end{equation}
where $\star$ indicates a convolution over momentum fractions, and we suppress all arguments other than the renormalization scales. The anomalous dimension $\bm{\Gamma}^H$ is a matrix in the space of parton multiplicities and an operator in color space. It consists of two parts, $\bm{\Gamma}^H=\bm{\Gamma}^S+\bm{\Gamma}^C$, describing soft-collinear and purely collinear dynamics, respectively. The purely collinear part $\bm{\Gamma}^C$ describes a modified DGLAP evolution of the PDFs above the scale $\mu_s$~\cite{Becher:2023mtx}. In leading-logarithmic approximation, however, its effect can be approximated by evaluating the PDFs at the high scale~$\mu_h$. The SLLs arise only from the soft-collinear part, which  may be decomposed further as~\cite{Becher:2021zkk}
\begin{equation}\label{eq:anomalous_dimension_soft_part}
   \bm\Gamma^{S} 
   = \gamma_{\rm cusp}(\alpha_s) \Bigl( \bm{\Gamma}^c
    \ln\frac{\mu^2}{\mu_h^2} +\bm{V}^G \Bigr)
    + \frac{\alpha_s}{4\pi}\,\overline{\bm{\Gamma}} + \mathcal{O}(\alpha_s^2) \,,
\end{equation} 
where $\gamma_{\rm cusp}$ is the cusp anomalous dimension. The piece enhanced by the logarithm generates double-logarithmic corrections in the cross section, but these terms only contribute in the presence of the Glauber phases contained in $\bm{V}^G$. The last term $\overline{\bm{\Gamma}}$ describes real and virtual soft emissions.

For the case of $2\to2$ kinematics, one can perform most of the integrals associated with the final-state partons in~\eqref{eq:factorization_formula} in a straightforward way. In the leading double-logarithmic approximation, this results in the triple-differential cross section
\begin{equation} \label{eq:d3sigma}
   \bigg(\frac{d^3\sigma}{dy_3\spac dy_4\spac dp_T}\bigg)_{\!\!2\to2} 
   = \frac{2\spac p_T}{s} \sum_{\substack{\text{partonic} \\ \text{channels}}}
    f_1(\xi_1,\mu_h)\,f_2(\xi_2,\mu_h)\,
    \bigg(\frac{d\hat{\sigma}}{dr}\bigg)_{\!\!12\to34}
\end{equation}
in the rapidities $y_3, y_4$ of the final-state partons and the transverse momentum $p_T$ of the two jets. Expressing the ratio $r=-\hat{t}/\hat{s}$ of partonic Mandelstam variables and the parton momentum fractions through these variables, one finds
\begin{equation} 
   r = \frac{1}{1+e^{\Delta Y}} \,, \qquad
   \xi_1 = \frac{p_T}{\sqrt{s}} \left( e^{y_3} + e^{y_4} \right) , \qquad
   \xi_2 = \frac{p_T}{\sqrt{s}} \left( e^{-y_3} + e^{-y_4} \right) ,
\end{equation}
where $\Delta Y\equiv y_3-y_4$. This implies $\hat s=\xi_1\spac\xi_2\spac s=2\spac p_T^2\left(1+\cosh\Delta Y\right)$. In~\eqref{eq:d3sigma} we sum over all partonic channels $1+2\to3+4$ and include the appropriate PDFs. The relevant partonic channels and their individual contributions to the cross section are listed in Table~\ref{tab:numerical_values} below. In writing~\eqref{eq:d3sigma}, one uses that the low-energy matrix element factorizes into a product of PDFs, see \eqref{eq:low_energy_ME}. In~\cite{Becher:2024kmk}, arguments were presented that suggest that PDF factorization is consistent with the RG evolution of the hard functions at least up to three-loop order.

\begin{figure}[t]
    \centering
    \includegraphics[scale=1]{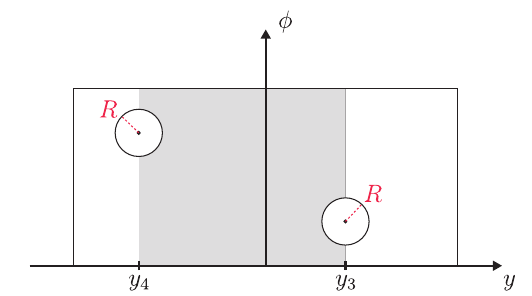}
    \caption{Definition of the gap region (gray) between the jets with radius $R$ in the rapidity and azimuthal plane used here and in the ATLAS analysis in~\cite{ATLAS:2011yyh}.}
    \label{fig:gap}
\end{figure}

To obtain the total gap-between-jets cross section, the differential cross section~\eqref{eq:d3sigma} is integrated over $p_T>p_{T\mathrm{cut}}$ and over the gap region of width $|\Delta Y|$. For our formalism to be applicable, the transverse momentum cutoff must be chosen such that a significant scale hierarchy between $p_{T\mathrm{cut}}$ and the veto scale $Q_0$ is ensured. We impose the jet veto on all emissions in the rapidity region between the two leading jets, as illustrated in Figure~\ref{fig:gap}. To ensure infrared safety at higher orders, the veto region does not include the hard jets. Our definition of the gap region agrees with the one used by the ATLAS collaboration in their measurement of the gap-between-jets cross section~\cite{ATLAS:2011yyh}. In this analysis, the jets are defined using anti-$k_T$ clustering, which is equivalent to a fixed-cone constraint at our accuracy level. An extension of the factorization formula to more general clustering algorithms was presented in~\cite{Becher:2023znt}. In our formalism, the dependence on the jet region is contained in the three angular integrals
\begin{equation}\label{eq:Jints}
   J_j \equiv \int\frac{d\Omega(n_k)}{4\pi} \left( W_{1j}^k - W_{2j}^k \right) 
    \Theta_{\rm veto}(n_k) \,, \qquad 
   W_{ij}^k = \frac{n_i\cdot n_j}{n_i\cdot n_k\,n_j\cdot n_k}
\end{equation}
with $j=2,3,4$ and $n_k$ restricted to be inside the gap region. For fixed jet radius $R$, we assume that the rapidity gap $\Delta Y$ is wide enough for the jets not to leak out on both sides ($|\Delta Y|>R$) and not to overlap ($\sqrt{|\Delta Y|^2+\pi^2}>2R$). We then obtain
\begin{equation} \label{eq:J12}
   J_{12} \equiv J_2 = |\Delta Y| - \frac{R^2}{2} \,, \qquad
   J_3 = - J_4 \,,
\end{equation}
and
\begin{equation} \label{eq:J4}
\begin{aligned}
   J_4 &= \Delta Y - \text{sgn}(\Delta Y)\,\frac{R}{\pi} 
    \int_0^1\!dx\,\bigg[\, \ln\frac{\cosh\left(R\sqrt{1-x^2}\right) - \cos R x}{1-\cos R x} \\
   &\hspace{4.9cm} + \ln\frac{\cosh\Delta Y + \cos R x}%
                             {\cosh\left(|\Delta Y|-R\sqrt{1-x^2}\right) + \cos R x} \,\bigg] \\
   &= \Delta Y - \frac{R^2}{4} \tanh\frac{\Delta Y}{2}
    - \text{sgn}(\Delta Y) \left[ \frac{2 R}{\pi} 
    + \frac{R^3}{6\pi} \left( \tanh^2\frac{\Delta Y}{2} - \frac23 \right)
    + \mathcal{O}(R^5) \right] .
\end{aligned}
\end{equation}
Note that the leading correction to $J_3$ and $J_4$ is linear in $R$ and not quadratic, as one would have naively expected. All higher-order terms contain only odd powers of $R$, but their numerical impact is negligible for $R<1$.

The Born-level partonic cross sections in~\eqref{eq:d3sigma} are given by
\begin{equation} \label{eq:partonic_Born}
    \bigg(\frac{d\hat{\sigma}}{dr}\bigg)_{\!\!12\to34} = \frac{1}{16\pi \spac \hat{s}} \, \big\langle \widetilde{\bm{\mathcal{H}}}_4(\mu_h) \, \bm{1} \big\rangle \,,
\end{equation}
where $\mu_h\sim \sqrt{\hat{s}}$. The tilde indicates the ``unintegrated'' hard functions, i.e.\ the hard-scattering amplitudes squared, which we need at lowest order in perturbation theory only. The resummed SLL contributions to the partonic cross sections can be expressed as a linear combination of eleven color structures $\bm{X}_i$ under the color trace with the hard functions, with coefficients depending on the scales $\mu_h$ and $\mu_s$~\cite{Becher:2023mtx,Boer:2024hzh}. The result can be written in the compact form
\begin{equation} \label{eq:partonic_SLL}
   \bigg(\frac{d\hat{\sigma}^\mathrm{SLL}}{dr}\bigg)_{\!\!12\to34} 
   = \frac{1}{16\pi\spac\hat{s}}\,
    \big\langle \widetilde{\bm{\mathcal{H}}}_4(\mu_h)\,\bm{X}^T \big\rangle\, 
    \mathbbm{U}_\mathrm{SLL}^{(2)}(\mu_h,\mu_s)\,\varsigma \,.
\end{equation}
The evolution matrix $\mathbbm{U}_\mathrm{SLL}^{(2)}(\mu_h,\mu_s)$, which acts on the space of the color structures $\bm{X}_i$, has been presented in (6.5) of~\cite{Boer:2024hzh} in RG-improved perturbation theory. The auxiliary vector $\varsigma^T=(1,0,\dots,0)$ projects out the first column of this matrix.

\begin{table}[t]
    \centering
    \begin{tabular}{|c|r|r||c|r|r|}
        \hline
        \cellcolor{Lightgray}process & \cellcolor{Lightgray}$\sigma_{2\to2} \; [\text{pb}]$ & \cellcolor{Lightgray}$\sigma_{2\to2}^\mathrm{SLL} \; [\text{pb}]$ & \cellcolor{Lightgray}process & \cellcolor{Lightgray}$\sigma_{2\to2} \; [\text{pb}]$ & \cellcolor{Lightgray}$\sigma_{2\to2}^\mathrm{SLL} \; [\text{pb}]$ \\ \hline\hline
        $\leftlap{q\bar{q}'}{qq}\to \rightlap{ q'\bar{q}'}{qq}$ & 231.5 & 12.0 & $q\bar{q} \to gg$ & 12.4 & $-$0.9 \\
        $\leftlap{q\bar{q}'}{qq'}\to \rightlap{ q'\bar{q}'}{qq'}$ & 454.4 & 22.2 & $qg \to qg$ & 4104.6 & 403.3 \\
        $\leftlap{q\bar{q}'}{q\bar q}\to \rightlap{ q'\bar{q}'}{q \bar q}$ & 142.0 & 7.4 & $gg \to q\bar{q}$ & 57.5 & $-$4.4 \\
        $\leftlap{q\bar{q}'}{q\bar{q}'}\to \rightlap{ q'\bar{q}'}{q\bar{q}'}$ & 372.9 & 18.0 & $gg \to gg$ & 2281.1 & 150.6 \\
        $\leftlap{q\bar{q}'}{q\bar q}\to q'\bar{q}'$ & 3.6 & $<$0.1 &&& \\
        \hline
        $\sum$ & 1204.4 & 59.6 & $\sum$ & 6455.6 & 548.6 \\
        \hline\hline
        \multicolumn{2}{|c|}{$\sum_{\text{all channels}}$ } & \multicolumn{2}{c|}{7660.0} & \multicolumn{2}{c|}{608.2} \\ \hline
    \end{tabular}
    \caption{Contributions of different partonic subprocesses to the integrated gap-between-jets cross section at the LHC for $\sqrt{s}=13$\,TeV, $p_T>200$\,GeV, $2<|\Delta Y|<3$, and jet radius $R=0.6$. The SLL contribution is shown for $Q_0=20$\,GeV. In the left portion of the table we present the channels involving only quarks and/or anti-quarks, while channels involving gluons are shown in the right portion. The (massless) quarks $q$ and $q'$ have different flavors and are summed over $u,d,s,c,b$ as well as the corresponding anti-quarks. The $qg\to qg$ contribution includes a factor~2 to account for the process $gq\to gq$.}
    \label{tab:numerical_values}
\end{table}

To calculate the traces of hard functions and color structures, it is convenient to introduce a basis $\{|\mathcal{C}_I\rangle\}$ of color configurations for a given partonic channel. These bases contain two elements for four-quark scattering, three for two-quark\,--\,two-gluon scattering, and nine for four-gluon scattering. The conventional choices are not necessarily orthogonal bases (see e.g.~\cite{Bern:2002tk}) and, therefore, the completeness relation reads
\begin{equation}
   \bm{1} = \sum_{I,J} \big| \mathcal{C}_I \big\rangle\, 
    \big(\bm{G}^{-1}\big)_{IJ} \, \big\langle\mathcal{C}_J \big| \,,
    \quad \text{with non-trivial Gram matrix} \quad
    \bm{G}_{IJ} = \big\langle\mathcal{C}_I \big| \mathcal{C}_J \big\rangle \,.
\end{equation}
Inserting this relation in~\eqref{eq:partonic_SLL} twice, and defining the matrix representations of the hard functions and color structures as
\begin{equation}
   \big(\widetilde{\bm{\mathcal{H}}}_4\big)_{IJ} 
   \equiv \sum_{K,L} \big(\bm{G}^{-1}\big)_{IK}\, 
    \big\langle\mathcal{C}_K \big| \widetilde{\bm{\mathcal{H}}}_4 \big| 
    \mathcal{C}_L \big\rangle \, \big(\bm{G}^{-1}\big)_{LJ} \,, \qquad
    \big(\bm{X}_i\big)_{IJ} \equiv \big\langle \mathcal{C}_I 
    \big| \spac\bm{X}_i \spac\big| \mathcal{C}_J \big\rangle \,,
\end{equation}
one can write the traces in the form
\begin{equation}
   \big\langle \widetilde{\bm{\mathcal{H}}}_4\,\bm{X}_i \big\rangle 
   = \frac{1}{\mathcal{N}_1 \spac \mathcal{N}_2} \sum_{I,J} 
    \sum_{\text{spins}} \big(\widetilde{\bm{\mathcal{H}}}_4\big)_{IJ}\, 
    \big(\bm{X}_i\big)_{JI} \,,
\end{equation}
where $\mathcal{N}_i=2N_c$ for parton $i$ being a quark or anti-quark, and $\mathcal{N}_i=(d-2)(N_c^2-1)$ for it being a gluon. One particular choice of color bases and the associated matrix representations for the spin-summed ``unintegrated'' hard functions have been given in~\cite{Broggio:2014hoa} for all relevant $2\to2$ processes up to NNLO. We have calculated the $\bm{X}_i$ matrices using \textsc{ColorMath}~\cite{Sjodahl:2012nk} and listed them in a supplemented \textsc{Mathematica} notebook.

Upon evaluating the color traces $\langle\ldots\rangle$ in~\eqref{eq:partonic_SLL}, we observe that for $q\bar{q}\to q\bar{q}$ scattering the SLL contribution to the $pp\to 2$\,jets cross section contains expansion coefficients $c_n$ of $\mathcal{O}(N_c)$ in \eqref{eq:SLLgeneric}, whereas for all other partonic channels these coefficients are of $\mathcal{O}(N_c^0)$. This leads to SLL contributions that are only suppressed by one power of $1/N_c$, an effect that to our knowledge has so far not been noticed in the literature. This enhancement can be traced back to the interference of two different color configurations in the amplitude. However, we find below that the $q\bar{q}\to q\bar{q}$ channel only contributes a small amount to the $pp\to2$\,jets cross section.

\section{Results}

\begin{figure}[t]
    \centering
    \includegraphics[scale=1]{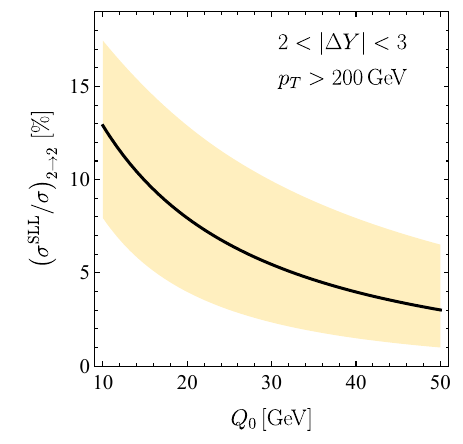}
    \caption{SLL contribution to the $pp\to2$\,jets cross section at the LHC as a function of the veto scale $Q_0$, for a center-of-mass energy $\sqrt{s}=13$\,TeV and jet radius $R=0.6$. The black curve shows the central result obtained in RG-improved perturbation theory. The perturbative uncertainties indicated by the yellow bands are obtained from the variation of the soft scale $\mu_s$ by a factor~2 about its default value~$Q_0$.}
    \label{fig:integrated_SLL}
\end{figure}

We are now in a position to determine the impact of the super-leading logarithms for the physical $pp\to 2$\,jets cross section. This involves integrals over the rapidities $y_3$, $y_4$, and transverse momentum $p_T$, which we evaluate numerically. We set the high scale to $\mu_h=2\spac p_T$, employ a jet radius $R=0.6$ and use the parton distribution functions from the NNPDF4.0 NLO set with $\alpha_s(M_Z)=0.118$ \cite{NNPDF:2021njg} via \textsc{ManeParse}~\cite{Clark:2016jgm}. As described above, the DGLAP evolution of the PDFs is taken into account by evaluating them at the high scale $\mu_h$. In Table~\ref{tab:numerical_values}, we present the individual contributions of the different partonic channels (including PDFs) to the integrated cross section for $p_T>200\,\mathrm{GeV}$, $\sqrt{s}=13\,\mathrm{TeV}$ and $2<|\Delta Y|<3$. For each channel, the first column shows the cross section at Born level and the second one the SLL contribution, obtained using the veto scale $\mu_s=Q_0=20\,\mathrm{GeV}$. It is evident that the main contributions both to the Born cross section ($\approx 84\%$) and even more so for the SLLs ($\approx 90\%$) are due to the gluonic channels $qg\to qg$ and $gg\to gg$.

The SLL contribution to the physical $pp\to 2$\,jets cross section, relative to the Born cross section, is shown in Figure~\ref{fig:integrated_SLL} as a function of the veto scale $Q_0$. It reaches up to $(13\spac_{-5}^{+4}\spac)\%$ for $Q_0=10\,\mathrm{GeV}$. However, as indicated by the yellow band, the scale uncertainties are still quite large in the leading double-logarithmic approximation. We have estimated them by varying the low scale $\mu_s$ by a factor~2 about the default value $Q_0$, while the high scale is kept fixed at $\mu_h=2\spac p_T$. For higher values of $Q_0$ the SLL contribution becomes smaller. However, taking the uncertainties into account, we find that it could still be as large as 10\% for $Q_0=30$\,GeV

\begin{figure}[t]
    \centering
    \includegraphics[scale=1]{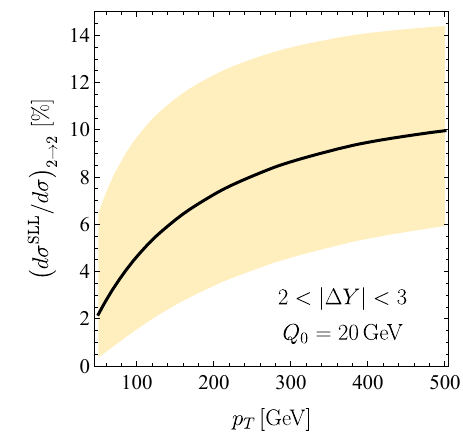}
    \includegraphics[scale=1]{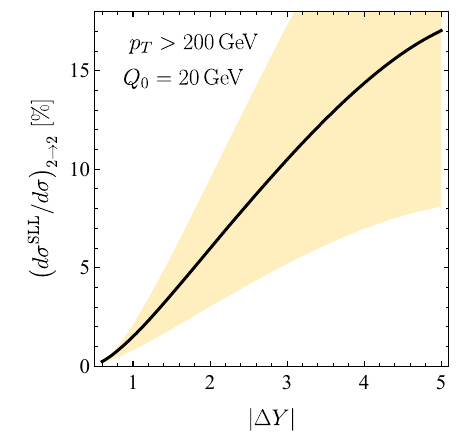}
    \caption{SLL contribution to the differential $pp\to2$\,jets cross section as function of transverse momentum $p_T$ (left) and gap size $|\Delta Y|$ (right). The veto scale is fixed to $Q_0=20$\,GeV.}
    \label{fig:differential_SLL}
\end{figure}

In Figure~\ref{fig:differential_SLL}, we present the contributions of the SLLs to the differential cross sections $d\sigma^{\mathrm{SLL}}/dp_T$ and $d\sigma^{\mathrm{SLL}}/d\Delta Y$, relative to the corresponding differential Born cross sections. In the first case, we observe that the SLL contribution increases with growing $p_T$, as the logarithm $L$ becomes larger. For small transverse momenta the contribution decreases, because the required scale hierarchy is no longer present. In the second case, one can see an approximately linear dependence on $\Delta Y$, in particular for $1<|\Delta Y|<3$. 
This is to be expected, as the angular integrals $J_j$ scale with $\Delta Y$, see \eqref{eq:J12} and \eqref{eq:J4}. For large $\Delta Y$, this dependence is altered due to the imposed cut on the transverse momentum.

\section{Outlook}

We have presented the first complete analysis of the contribution of super-leading logarithms (SLLs) to a physical gap-between-jets cross section at the LHC, including all partonic channels and the corresponding parton distribution functions, as well as quantum interference effects in the squared partonic scattering amplitudes. Earlier papers have analyzed individual partonic channels and found sizable both positive and negative contributions \cite{Becher:2021zkk,Becher:2023mtx,Boer:2024hzh}, so one could speculate whether cancellations would reduce the magnitude of the effect in the hadronic cross section.  Our analysis shows that this is not the case; the contribution of super-leading logarithms remains relevant and is of similar size at the partonic and hadronic levels. We have presented numerical results for integrated and differential $pp\to 2$\,jets cross sections, finding large corrections for low $Q_0$ values, albeit with significant scale uncertainties. Importantly, our analysis also includes a three-loop contribution resulting from two Glauber phases, which has the same physical origin as the SLLs. Note that this three-loop contribution by itself would overestimate the effect by about a factor~2.

On a more theoretical level, our analysis has also revealed the surprising result that for $q\bar q\to q\bar q$ scattering the SLL contribution is only suppressed by one power of $1/N_c$, as opposed to $1/N_c^2$ for all other $2\to 2$ subprocesses. This arises due to interference effects in the partonic amplitudes, which were not included in the earlier literature. 

Our work represents a first step in the systematic resummation of non-global hadron-collider observables and demonstrates the potential numerical importance of SLL effects. At the same time, a lot of interesting physics remains to be explored before precise predictions for such observables will be obtained. Our resummation includes the effect of the running coupling, but there are several other next-to-leading logarithmic effects which are not yet accounted for. They are generated by higher insertions of non-logarithmic terms in the anomalous dimension  $\bm{\Gamma}^H$, which fall in three categories: Glauber phases $\bm{V}^G$, wide-angle soft emissions $\overline{\bm{\Gamma}}$, and collinear terms $\bm{\Gamma}^C$. The contributions of multiple insertions of $\bm{V}^G$ have been calculated in~\cite{Boer:2023jsy,Boer:2023ljq} and build up the so-called ``Glauber series'', which remarkably can be summed to all orders in the large-$N_c$ approximation~\cite{Boer:2024xzy}, even though the effect itself is color suppressed. Insertions of $\bm{\Gamma}^C$ have been accounted for only approximately in the present work by evaluating the PDFs at the high scale $\mu_h$. It would be very interesting to explore the effects of non-DGLAP evolution above the scale $\mu_s=Q_0$ in future work.

Finally, and most importantly, one should properly take into account the soft emission operator $\overline{\bm{\Gamma}}$, higher insertions of which generate the non-global logarithms (NGLs). The resummation of these logarithms for the ATLAS gap-between-jets measurement \cite{ATLAS:2011yyh} has been studied in \cite{Hatta:2013qj,Balsiger:2018ezi} working in the strict $N_c\to\infty$ limit, in which SLLs and Glauber phases are absent. As one would expect, the higher-order emissions have the effect of suppressing the cross section at small $Q_0$. For $Q_0$ in the interval between 20 and 40\,GeV, for example, the reduction is about a factor of~2. In the meantime, also methods for including subleading NGLs in the resummation have been developed~\cite{Banfi:2021owj,Banfi:2021xzn,Becher:2021urs,Becher:2023vrh}, which help to reduce the large scale uncertainties. However, it is currently unknown how to combine the NGLs with the SLLs. A simple multiplication of the two effects is likely to be insufficient to capture their intricate interplay. A complete treatment of NGLs would require their computation at finite $N_c$ in the presence of Glauber phases. In light of the ongoing progress in finite-$N_c$ computations~\cite{Hatta:2013iba,Hatta:2020wre,AngelesMartinez:2018cfz,Nagy:2019pjp,Forshaw:2019ver,Boer:2024xzy}, this might indeed become feasible in the future. Alternatively, it is conceivable that large-$N_c$ methods can help to understand the interplay of NGLs with SLLs at least in an approximate way.

\subsection*{Acknowledgments}

We thank J\"urg Haag and Nicolas Schalch for comments on the manuscript. PH thanks the Munich Institute for Astro-, Particle and BioPhysics (MIAPbP) for hospitality during the final stages of this work, and MN gratefully acknowledges support from the Albert Einstein Center for Fundamental Physics (AEC) at the University of Bern. This research has received funding from the European Research Council (ERC) under the European Union’s Horizon 2022 Research and Innovation Program (ERC Advanced Grant agreement No.~101097780, EFT4jets) and from the Swiss National Science Foundation (SNF) under grant 200021\_219377. The work reported here was also supported by the Clusters of Excellence \emph{PRISMA${}^+$} (EXC 2118/1, Project ID 390831469) and \emph{ORIGINS} (EXC 2094, Project ID 390783311) funded by the German Research Foundation (DFG) within the Germany Excellence Strategy. Views and opinions expressed in this work are those of the authors only and do not necessarily reflect those of the European Union or the European Research Council Executive Agency. Neither the European Union nor the granting authority can be held responsible for them. 

\clearpage
\pdfbookmark[1]{References}{Refs}
\bibliography{refs.bib}

\end{document}